# Full Field Transmission Tomography (FFOTT) for imaging extrachromosomal circular DNA (eccDNA) in cancer cell nuclei

Nathan Boccara, Samer Alhaddad, Viacheslav Mazlin

Institut Langevin, ESPCI Paris, PSL University, CNRS. 1 rue Jussieu 75005 Paris France.

**Abstract**
   Detecting the specificity of cancer cells to distinguish them from normal ones is an important step in the general framework of cancer diagnosis. A routine example of such diagnosis in cancerous tissues implies using microscope analysis of fixed, paraffined, and colored slices such as the H&E stain (1). Such a method, which takes place after surgery, is based on carefully analyzing the cell's size and shape. Often, this approach is performed in parallel with more modern genetic tests. Recent research has hypothesized that extrachromosomal circular DNA (eccDNA) could be considered a new hallmark of cancer (4). Thus, this research aims to check if using a simple, label-free microscope dynamic analysis performed on living cancer cells would allow efficient and simpler detection of cancer cells.

**Introduction**
   Extrachromosomal DNA (ecDNA) plays a significant role in cancer biology, particularly in oncogene amplification and tumorigenesis. Originally identified over 50 years ago as minute chromatin bodies and later called double minutes, ecDNAs are large, highly amplified, circular DNA structures that separate from chromosomes. Advances in ultra-structural microscopy, whole genome sequencing, computational reconstruction, and long-range DNA optical mapping have confirmed the circular nature of ecDNAs. These structures serve as templates for RNA transcription, driving high levels of gene expression and contributing to the genetic heterogeneity of tumors. Oncogenes, which are mutated forms of proto-oncogenes, become problematic when their copies are excessively amplified, a common molecular alteration in cancer. Recent research indicates that oncogene-carrying ecDNA formation significantly contributes to gene amplification and the resultant genetic diversity within tumors. This process involves double-strand breaks and chromosomal rearrangements, leading to the circularization of DNA segments and subsequent replication in the absence of tumor suppressor responses, promoting a rapid increase in ecDNA and oncogene copies (2).
EcDNAs are frequently found in cancer cells, prompting renewed interest in their role in oncogene amplification. The understanding of ecDNA biology has evolved beyond descriptive studies, thanks to foundational work from Schimke's lab, which proposed that ecDNAs are unstable forms of gene amplification. These particles differ from chromosomal DNA in several key biological aspects, including their replication, transcription, and repair mechanisms. To further investigate ecDNA's role in cancer, new computational and cell biology tools are necessary. Techniques like CRISPR-Cas and advanced cell imaging are crucial for deciphering the mechanisms behind ecDNA formation, maintenance, and its impact on cancer pathogenesis. By promoting rapid genomic diversification, ecDNAs enhance the ability of cancer cells to adapt to changing environments, including therapeutic pressures, highlighting the importance of understanding and targeting ecDNA in developing effective cancer treatments (3).
Extrachromosomal circular DNA (eccDNA) is a type of circular DNA that has attracted the attention of researchers and is now considered a hallmark in cancer (4). The oncogenes seem to be over-expressed by eccDNA and induce the growth of malignant tumors. Nevertheless, the exact mechanism of these interactions has yet to be established.

The sizes distribution of eccDNA of seven cell lines was found in the 1 to 10 kb (6); few images are available mostly from electron microscope although confocal microscopy shows ring structure in the half micrometer range (7) e.g. in the liver the diameters of eccDNA expand from a few tens of base pairs up to a few thousands base pairs.

Recently the Full Field Transmission Tomography (FFOTT) approach was proposed by Mazlin et al. (9). FFOTT is a microscope-based interferometric technique that takes advantage of the $\pi$ Gouy phase shift that happens close to the focus of the objective to induce phase variations in the interferences between the illuminating light and the scattered one (9). FFOTT has been applied successfully to algae physiology(11)

This letter aims to start exploring the use of Full Field Optical Transmission Tomographic microscopy, knowing that this method will be difficult to use when the diameter of the structure under examination is smaller than 400 nm. This drawback is balanced by a unique feature of FFOTT that allows it to work with living cells and follow their dynamic behavior.

**Setup and signal processing.**
A schematic diagram is shown in Figure 1; The FFOTT setup is particularly straightforward: the source is a LED (455 nm Royal blue, Thorlabs mounted LED) the objective is a 100X, 1.25 NA (Amscope Plan Achromatic microscope objective). The camera is a black and white 1 Megapixel camera (Photon Focus MV-D1024E series 150 f/s). The resolution r is given by: $R=0.61 l/NA=0.22$ μm and the depth of field by $D=1.22 n\lambda/NA^2=0.54$ μm ($\lambda$ is the wavelength and n the refractive index). If we consider the circular shape of the eccDNA structure it looks difficult to image structures whose diameter is <0.4 μm.

The field of view was either 100x100 μm$^2$ or 60x60 μm$^2$ depending on the magnification linked to the tube lens focal length.

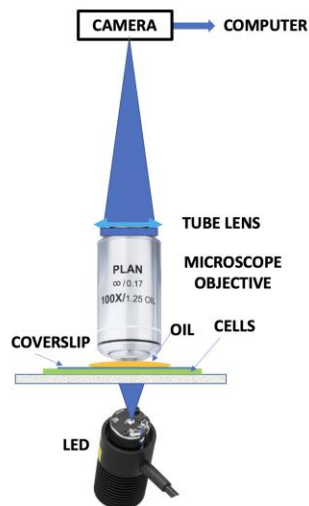

Figure 1: experimental setup used for the FFOTT experiment (see text for detailed information).

As discussed by Mazlin (1), FFOTT takes its origin through the interference of the illuminating incident beam and the beam scattered by the structures close to the focus of the microscope objective.

Figure 2 summarizes this interference behavior using a vectorial representation of waves in 3 cases along the depth of field: before and after the focus and in between with still and fluctuating positions. The sectioning that is associated to the movement of intracellular structures requires only recording for a few seconds a number of successive images (typically 100), the movement induces a phase shift and correlatively an intensity variation. We will consider here only this mode that does not require any mechanical displacement of the object or of the objective but uses intracellular structures.

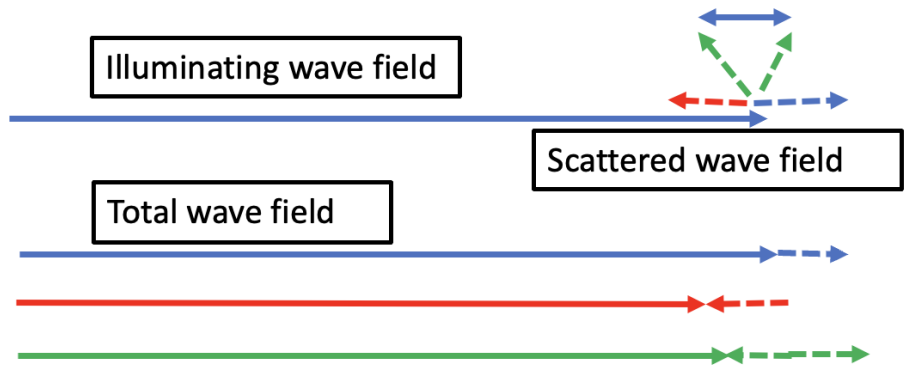

Figure 2. Vectorial representation of the optical fields that are involved in FFOTT as a function of the position of a scattering object about the objective focus: In blue and red constructive and destructive interferences lead to an increase or a decrease in the total amplitude respectively. In green the intermediate situation where the object fluctuates in position around the focus

We typically record a stack of 100 images to reveal the dynamic signal associated to the movements of sub-nuclear structures. We used ImageJ free software (11) to display the standard deviation (STD) of each pixel of the stack. This very simple processing works well but more elaborate processing could be achieved using Matlab or Python written programs.

**Experimental Results**

We first used the dynamic FF-OTT method by studying the behavior of HeLa cells (93021013-1VL, Merck) grown on a cover slip in a Petri dish. The cells were positioned in the best focus of the FF-OTT device. Here we used a high-resolution 100X oil immersion FFOTT objective mentioned above.

To stimulate the nuclei dynamic activity we analyzed the data taken after adding Desoxyglucose (an inhibitor of glycolysis) and observed several ring structures underlined by the arrow in Figure 3. Such rings can be only observed when the ring-shaped structures appear in frontal view; when these structure's planes are tilted only part of them are visible.

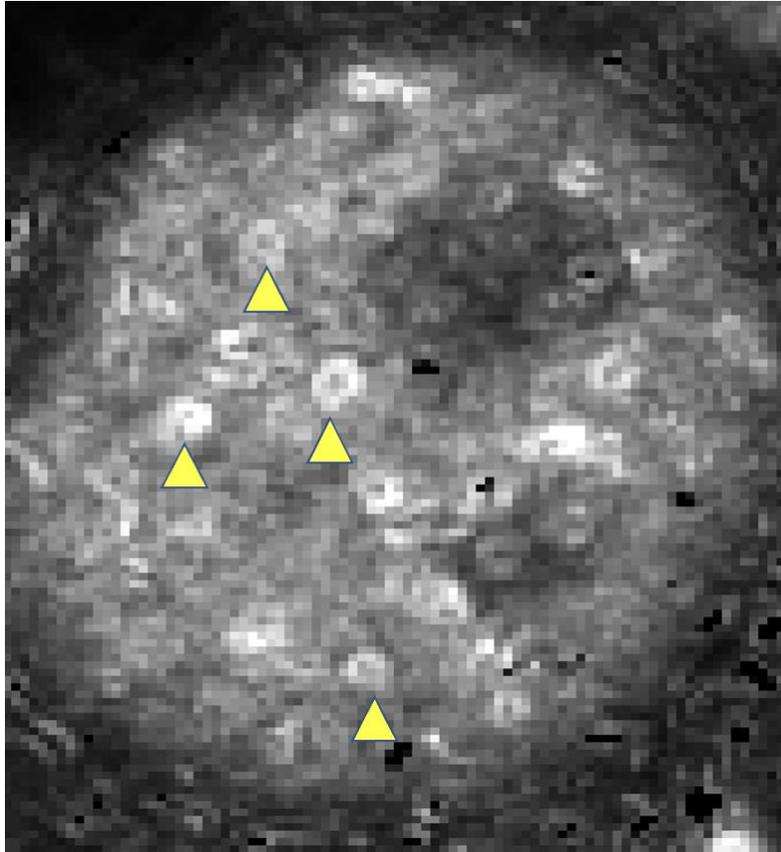

*Figure 3: Ring structures underlined by yellow triangle observed in HeLa nuclei.(image size 6x6 µm$^2$)*

The African green monkey kidney cells (COS-7) were also plated on coverslips without any treatment. The dynamic signal was strong enough to reveal circular structures with much less background than in HeLa cells. These structures appear slightly larger than with HeLa cells (0.6 µm instead of 0.5 µm), figure 4.

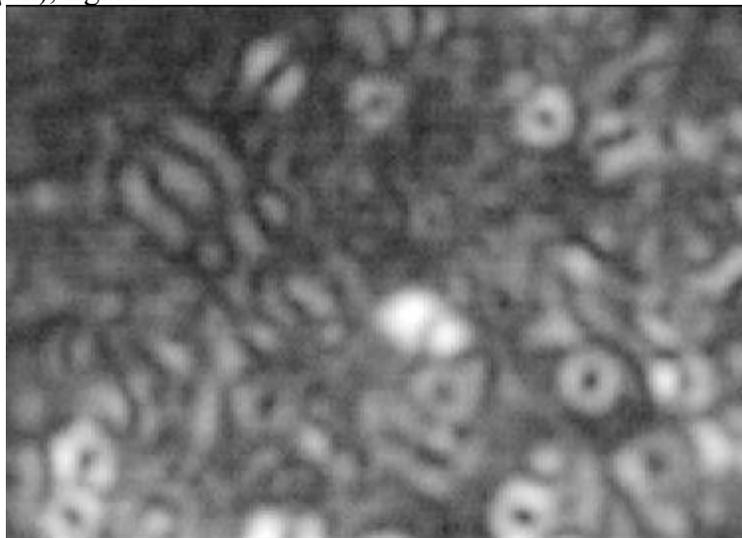

*Figure 4. Ring structures observed in COS cell nuclei. (image size 10x7.5 µm$^2$)*

Finally, Lovo cells (epithelial cells from a human colon tumor) one can see on figure 5 ring structures with two main different characteristics: the signal was found weaker and the ring diameter size slightly larger larger (around 2 µm).

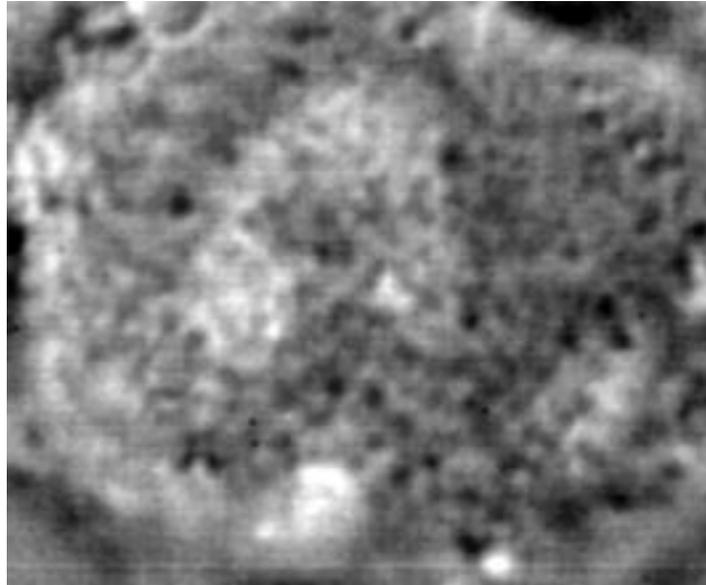

*Figure 4. Ring structures observed in LOVO cell nuclei. (image size 16x13 µm$^2$)*

To display the round sub-micrometer structures of figure 3, 4 and 5, we zoomed the recorded images over the size of a nucleus but our field of view can handle much more cells.

**Conclusion**

As mentioned in the abstract we do not claim that the circular structures that we observed in nuclei of cancer cells should be fully assigned to eccDNA; indeed much more bioanalysis is required, but if the structures observed here happen to be eccDNA this opens the path to further use of FFOTT in cancerology. Indeed, we have tentatively assigned eccDNA through their shape, but a more complete analysis of the dynamic behavior should bring supplementary information. Improvement could be made in the image analysis using free or paid computing tools such as the ones that allow counting specific ring shapes and their density.

**Acknowledgments**
N.B. would like to acknowledge Doctor Andrea Ventura from MSK (New York, USA) who sent him illuminating documents about ecDNA, and to thank the Langevin Institute's scientists for their patience and enthusiasm.

## REFERENCES

1) *H&E Staining Overview: A Guide to Best Practices* https://www.leicabiosystems.com/en-fr/knowledge-pathway/he-staining-overview-a-guide-to-best-practices/

2) Wu, S., Bafna, V., & Mischel, P. S. (2021). Extrachromosomal DNA (ecDNA) in cancer pathogenesis. *Current opinion in genetics & development*, *66*, 78–82. https://doi.org/10.1016/j.gde.2021.01.001

3) Verhaak, R. G. W., Bafna, V., & Mischel, P. S. (2019). Extrachromosomal oncogene amplification in tumour pathogenesis and evolution. *Nature reviews. Cancer*, *19*(5), 283–288. https://doi.org/10.1038/s41568-019-0128-6
4) Li, Z., & Qian, D. (2024). Extrachromosomal circular DNA (eccDNA): from carcinogenesis to drug resistance. Clinical and experimental medicine, 24(1), 83. https://doi.org/10.1007/s10238-024-01348-6
5) Cross R. *The curious DNA circles that make treating cancer so hard:* chemical & engineering news volume 98, 40, (2022)
6) Richard P. Koche et al. *Extrachromosomal circular DNA drives oncogenic genome remodeling in neuroblastoma* Nat Genet. January; 52(1): 29–34 (2020)
7) Sihan Wu et al., *Circular ecDNA promotes accessible chromatin and high oncogene expression*, Nature, 575, 699–703 (2019)
8) Henssen A, *How circular DNA causes cancer in children* https://www.mdc-berlin.de/news December 16, 2019.
9) Mazlin et al. *Label free optical transmission tomography for biosystems: intracellular structures and dynamics*. Biomedical Optics Express Vol. 13, 8, 4190. (2022)
10) Giancoli. Physics. Chapter 25. *Optical Instruments*. Pearson Education, Inc (2014).
11) Bey et al. Dynalic Celle Imaging : application to the diatom Phaeodactylum tricornutum under environmental stresses, European Journal of Phycology, 58:2, 145-155.
12) Rasband, W.S., ImageJ, U. S. National Institutes of Health, Bethesda, Maryland, USA, https://imagej.net/ij/, 1997-2018
13) Giancoli. Physics. Chapter 24. *The wave nature of light*. Pearson Education, Inc (2014).
14) Tsang, V. T. C., Li, X., & Wong, T. T. W. (2020). A Review of Endogenous and Exogenous Contrast Agents Used in Photoacoustic Tomography with Different Sensing Configurations. *Sensors (Basel, Switzerland)*, *20*(19), 5595.